\RequirePackage{arydshln}
\documentclass[aps,twocolumn,nofootinbib,superscriptaddress,preprintnumbers,pra,10pt]{revtex4-1}

\usepackage{float}
\usepackage{colortbl}
\usepackage{amsmath,amssymb}
\usepackage{dsfont} 
\usepackage{hyperref}
\usepackage{graphicx}
\usepackage{enumitem}
\usepackage{arydshln}
\usepackage{mathtools}
\usepackage{bbold}
\usepackage{soul}
\makeatletter
\g@addto@macro\bfseries{\boldmath}
\makeatother

\def\deriv {\ensuremath{\mathrm{d}}}

\newcommand{\be} {\begin{equation}}
\newcommand{\ee} {\end{equation}}
\newcommand{\bea} {\begin{eqnarray}}
\newcommand{\eea} {\end{eqnarray}}
\newcommand{\no} {\nonumber}
\newcommand{\cO}{{\mathcal O}}
\newcommand{\cL}{{\mathcal L}}
\newcommand{\cA}{{\mathcal A}}
\newcommand{\cB}{{\mathcal B}} 
\newcommand{\cM}{{\mathcal M}} 
\renewcommand{\Re}{{\rm Re}}
\def\bsll{{\ensuremath{b\to s \ell^{+}\ell^{-}}}\,}

\newcommand{\rpk}{{$R_{pK}$}} 
\newcommand{\rkpi}{{$R_{K\pi}$}} 
\newcommand{\rkpipi}{{$R_{K\pi\pi}$}} 
\newcommand{\Rx}{{$R_{X}$}} 

\newcommand{\etasevenpk}{\eta^{77}_{pK}}
\newcommand{\etasevenkpipi}{\eta^{77}_{K\pi,K\pi\pi}}
\newcommand{\etanot}{\eta^{0}_X}
\newcommand{\etaseven}{\eta^{77}_X}
\newcommand{\etaone}{\eta^{79}_X}
\newcommand{\etatwo}{\eta^{79\prime}_X}
\newcommand{\etai}{\eta^{i}_X}
\newcommand{\etaelli}{\eta^{i,\ell}_X}
\newcommand{\rnot}{{\left< \etanot  \right> }} 
\newcommand{\rseven}{{\left<  \etaseven \right> }} 
\newcommand{\rseveni}{{\left<  \eta^{7i}_X \right> }} 
\newcommand{\rone}{{\left< \etaone \right> }}
\newcommand{\rtwo}{{\left< \etatwo \right> }} 
\newcommand{\ri}{{\left< \etai \right>}} 
\newcommand{\ril}{{\left< \etaelli \right>}} 
\newcommand{\rsevenpk}{{\left<  \etasevenpk \right> }} 
\newcommand{\rsevenkpipi}{{\left<  \etasevenkpipi \right> }} 

\newcommand{\kstar}{{$K^{*}(892)^{0}$\,}}

\begin{document}


\title{A general effective field theory description of \bsll lepton universality ratios}

\author{Gino Isidori}

\author{Davide Lancierini}

\author{Abhijit Mathad}

\author{Patrick Owen}

\author{Nicola Serra}

\author{Rafael Silva Coutinho}
\affiliation{Physik-Institut, Universit\"at Zu\"rich, CH-8057 Z\"urich, Switzerland}

\begin{abstract}
\vspace{5mm}
We construct an expression for a general lepton flavour universality (LFU) ratio, $R_{X}$, 
in \mbox{\bsll} decays 
in terms of a series of  hadronic quantities which can be treated as nuisance parameters. 
This expression allows to include any LFU ratio in global fits of \bsll short-distance parameters, even in the absence of a precise knowledge of the corresponding hadronic structure. The absence of sizeable LFU violation and the approximate left-handed structure of the Standard Model amplitude imply that only a very limited set of hadronic parameters 
hamper the sensitivity of $R_X$ to a possible LFU violation of short-distance origin.
A global \bsll combination is performed including the measurement of \rpk\  for the first time, resulting in a significance of new physics of $4.2\,\sigma$. 
In light of this, we evaluate the impact on the global significance of new physics using a set of experimentally promising non-exclusive $R_X$ measurements that LHCb can perform, and find that they can significantly increase the discovery potential of the experiment. 
\vspace{3mm}
\end{abstract}

\maketitle
\allowdisplaybreaks

\section{Introduction}\label{sec:intro}

In recent years, a pattern of deviations with respect to Standard Model (SM) predictions has manifested in measurements of \bsll processes. These include deviations in the angular distribution of the decay $B^0\to K^{*0}\mu^+\mu^-$~\cite{Aaij:2013qta,Aaij:2015oid,Aaij:2020nrf,Aaij:2020ruw}, a deficit in the decay rates~\cite{Sirunyan:2019xdu,Aaboud:2018mst,Aaij:2014pli,Aaij:2016flj,Aaij:2021pkz,LHCb:2021awg,LHCb:2021vsc} and deviations in lepton flavour universality (LFU) ratios~\cite{Aaij:2014ora,Aaij:2017vbb,Aaij:2019wad,Aaij:2021vac}. Within the framework of effective field theories, these deviations are 
numerically consistent with each other, pointing to a well-defined  hypothesis of new physics 
of short-distance origin~\cite{Ciuchini:2020gvn,Alguero:2021anc,Altmannshofer:2021qrr,Geng:2021nhg,Hurth:2021nsi,Cornella:2021sby}. 
Even under highly conservative theoretical assumptions, the global significance of the new physics 
hypothesis is as large as $4.3\sigma$~\cite{Isidori:2021vtc}.


Among these deviations, the LFU ratios are particularly interesting as their SM uncertainty is very precise~\cite{Hiller:2003js,Bordone:2016gaq,Isidori:2020acz}. They are defined within a region of squared dilepton invariant mass ($q^{2}$) as
\begin{equation}
    R_{X} \equiv  \dfrac{\displaystyle\int_{q^2_\mathrm{min}}^{q^2_{\rm max}} \dfrac{\deriv \Gamma (H_b\to X_s \mu^+\mu^-)}{\deriv q^2} \deriv q^2}{\displaystyle\int_{q^2_{\rm min}}^{q^2_{\rm max}} \dfrac{\deriv \Gamma (H_b\to X_s e^+e^-)}{\deriv q^2} \deriv q^2} ~.
    \label{eq:RX}
\end{equation}
where $H_b$ represents a $b$-hadron 
(meson or baryon)
and $X_s$ represents a well-defined hadronic system 
with strangeness, such that the 
transition satisfies $\Delta B=\Delta S$.

While the SM prediction $R^{\rm SM}_X=1$ is very robust,\footnote{We assume the 
$q^2$ range extends 
well above the dilepton mass threshold.}
 the precise cancellation of hadronic uncertainties can be broken in 
presence of new physics (NP). Namely, the interpretation of a new physics structure affecting these LFU ratios relies on 
the knowledge of the hadronic structure of the decays involved. This is why the LFU ratio \rpk~\cite{Aaij:2019wad} has not been included yet in \bsll  global fits, despite its clean SM prediction.
The same problem holds for any LFU ratio which contains a mixture of overlapping/interfering hadronic resonances where the underlying structure is unknown, referred to in the following as non-exclusive \Rx  ratios. Examples of this type are the LFU ratios \rkpipi\ and \rkpi, where for the latter the $K\pi$ system has an invariant mass larger than the \kstar resonance. The experimental prospects for these ratios are promising but their interpretation 
in terms of \bsll short-distance dynamics is not obvious.

Here, we propose a new method that allows to interpret any LFU ratio within the framework of effective Lagrangians for the first time, even if the detailed structure of the hadronic matrix elements is unknown.
The key observation that allows us to reduce the number of unknown handronic quantities is the fact that the 
SM amplitude is both lepton flavour universal and approximately left-handed. These two properties 
imply that  only a very limited set of NP amplitudes can yield sizeable non-standard contributions to $R_X$.
Their contribution can be described in terms of very few combinations of hadronic parameters, 
which can in turn be treated as nuisance  parameters.

The theoretical decomposition of $R_X$ following this logic is presented in Sect.~\ref{sec:theory}. 
Using this decomposition we perform a  global \bsll combination including the measurement of \rpk\ for the first time,
improving upon the global estimate of the significance presented in Ref.~\cite{Isidori:2021vtc}.
Using this method we also explore the potential impact of the expected measurements of \rpk, \rkpipi, and \rkpi\ 
with the full dataset collected so far by LHCb (Sec.~\ref{sec:prospect}). 
The conclusions of our analysis are summarised in Sect.~\ref{sec:conclusions}.

\section{General expression of $R_X$ in terms of Wilson coefficients}\label{sec:theory}

In the limit of heavy new physics, we can describe both SM and NP effects in $b\to s\ell^+\ell^-$ decays
by means of an effective Lagrangian containing only light SM fields. 
We normalise it as 
\be
\Delta \cL^{b\to s\ell\ell}_{\rm eff} = \frac{4 G_F}{\sqrt{2}} \frac{\alpha}{4\pi} V_{ts}^* V^{\phantom{*}}_{tb}
\,   
 \sum_{i } C_i \cO_i + {\rm h.c.}\,,
\ee 
where $G_F$ and $\alpha$ 
denote the Fermi constant and the 
electromagnetic coupling, respectively, 
and $V_{ij}$ denotes the elements of the Cabibbo-Kobayashi-Maskawa matrix.
The only difference between the SM and NP
cases lies in the number of effective operators, which is larger in a generic NP framework.
In full generality
the dimension-six operators with a non-vanishing tree-level matrix element in 
 $b\to s\ell^+\ell^-$ decays can be composed into three sets: i) dipole operators,
\be
\cO_{7}= \frac{m_b}{e} (\bar{s}_L\sigma_{\mu\nu} b_R)F^{\mu\nu}\, , \qquad  \cO^\prime_{7}= 
\frac{m_b}{e}
(\bar{s}_R\sigma_{\mu\nu} b_L)F^{\mu\nu} \, ,
\label{eq:dipole}
\ee
ii) vector operators,
\be
\!\!
\begin{array}{ll}
\cO^\ell_{9}=  (\bar{s}_L\gamma_\mu b_L)(\bar\ell \gamma^\mu\ell)  \,,  \quad 
& \cO^\ell_{10}=  (\bar{s}_L\gamma_\mu b_L)(\bar\ell\gamma^\mu\gamma_5\ell) \, , \\
 \cO^{\ell\prime}_{9}=  (\bar{s}_R\gamma_\mu b_R)(\bar\ell \gamma^\mu\ell)  \,,  \quad 
& \cO^{\ell\prime}_{10}=  (\bar{s}_R\gamma_\mu b_R)(\bar\ell\gamma^\mu\gamma_5\ell)\, , 
\end{array}
\label{eq:Olist}
\ee
and ii) scalar operators,
\be
\cO^\ell_{\hat S}=  (\bar{s}_L  b_R)(\bar\ell_R \ell_L)  \,,  \qquad\
\cO^{\ell\prime}_{\hat S}=  (\bar{s}_R  b_L)(\bar\ell_L  \ell_R) \, .
\label{eq:scalars}
\ee
In the NP case  the $\ell=e$ and~$\ell=\mu$ terms should be treated separately.
The scalar operators lead to \bsll  amplitudes which are helicity suppressed and 
can be safely neglected in most of the observables we are interested in. 
The only exception being the (single) effective combination which contributes 
to the $B_s\to \mu^+\mu^-$ helicity-suppressed rate. 
The dipole operator $\cO^\prime_7$ is negligible in the SM
and is severely constrained by 
$\Gamma(B\to K^*\gamma)$ and $\Gamma(B\to K^*\ell^+\ell^-)$ at low $q^2$~\cite{LHCb:2020dof}.
To describe SM and NP effects in the $R_X$ ratios, 
we can thus limit our  attention to the SM dipole operator ($\cO_7$) and the four vector operators in 
Eq.~(\ref{eq:Olist}).

As pointed in~\cite{Hiller:2014ula}, 
to elucidate general properties of the LFU ratios beyond the SM,
irrespective of the detailed structure of the hadronic matrix elements,
it is convenient to write the decay amplitudes in a 
basis of chirally projected operators. To do so, we introduce the combinations
\bea
C^{\ell}_L = C^{\ell}_9 - C^{\ell}_{10}\,, \qquad C^{\ell\prime}_L = C^{\ell\prime}_9 - C^{\ell\prime}_{10}\,, \no \\
C^{\ell}_R = C^{\ell}_9 + C^{\ell}_{10}\,, \qquad C^{\ell\prime}_R = C^{\ell\prime}_9 + C^{\ell\prime}_{10}\,.
\eea
With this notation,
the generic $H_b \to X_{s} \ell^+\ell^-$ transition amplitude can be decomposed as
\be
\cA(H_b \to X_s \ell^+\ell^- )  \propto  (\cM_{X,L}^\ell)^\alpha  (J^\ell_L)_\alpha  +   (\cM_{X,R}^\ell)^\alpha    (J^\ell_R)_\alpha 
\label{eq:Ampdec}
\ee
where 
\be
(J^\ell_L)_\alpha  = \bar{\ell}_L\gamma^\alpha \ell_L\,,   \qquad 
(J^\ell_R)_\alpha  = \bar{\ell}_R\gamma^\alpha \ell_R\,, 
\ee
and 
\bea
(\cM_{X,L}^\ell)^\alpha  & =  & C^{\ell}_L  J_X^\alpha +  C^{\ell\prime}_L  J^{\prime \alpha}_X +  C_7  J^{7\alpha}_X   \no\\ 
(\cM_{X,R}^\ell)^\alpha  & =  &  C^{\ell}_R  J_X^\alpha +  C^{\ell\prime}_R  J^{\prime \alpha}_X +  C_7  J^{7\alpha}_X  
\eea
with
\bea
&& J_X^\alpha = \langle X_s | \bar{s}_L\gamma^\alpha b_L | H_b \rangle\, ,  \qquad  
J^{\prime \alpha}_X = \langle X_s | \bar{s}_R\gamma^\alpha b_R | H_b \rangle\, ,   \no \\
&& J_X^{7\alpha}   \propto \frac{1}{q^2} q_\nu   \langle X_s | \bar{s}_L \sigma^{\alpha\nu} b_R | H_b \rangle\, .
\eea
In the limit where we neglect small lepton mass effects, the terms in Eq.~(\ref{eq:Ampdec}) proportional to 
the left-handed and right-handed leptonic currents do not interfere. Moreover, the following relation holds
\be
\left| \cM_{X,R}^\ell \right|^2   = \left| \cM_{X,L}^\ell \right|^2_{ \{ C^\ell_L \to C^\ell_R,\  C^{\ell\prime}_L \to C^{\ell\prime}_R \} }\,.
\ee
Integrating over all kinematic variables 
but for $q^2$, we can thus decompose the decay rate as
\be
\frac{d\Gamma_X^{\ell}}{dq^2}   = \frac{d\Gamma^{\ell}_{X,L}}{dq^2} + \frac{d\Gamma^{\ell}_{X,R}}{dq^2}\, ,
\label{eq:RXth1}
\ee
with
\be
\frac{d\Gamma^{\ell}_{X,R}}{dq^2} =  \left. \frac{d\Gamma^{\ell}_{X,L}}{dq^2}  \right|_{ \{ C^\ell_L \to C^\ell_R,\  C^{\ell\prime}_L \to C^{\ell\prime}_R \} } .
\label{eq:RXth2}
\ee
The explicit expression of   $d\Gamma^{\ell}_{X,L}/dq^2$ in terms of Wilson coefficients is 
\bea
&& \frac{d\Gamma^{\ell}_{X,L}}{dq^2}  = f_X^{\ell}(q^2) \Big\{ \left|C^{\ell}_L\right|^2 + \left| C^{\ell\prime}_L\right|^2  
  + \Re\Big[ 
 \etanot (q^2)  C^{\ell*}_L C^{\ell\prime}_L  \Big]
 \no\\ 
 && + \etaseven(q^2) |C_7|^2  
  + \Re\Big[ \etaone(q^2) C^*_7   C^{\ell}_L   +\etatwo (q^2) C^*_7  C^{\ell\prime}_L   \Big] \Big\}\,,  
  \no\\
\label{eq:rx_dGdq}
\eea
where $f_X^{\ell}(q^2)$ and the four $\etai(q^2)$ are channel-dependent hadronic parameters. The hadronic matrix elements 
$J_X^\alpha$ and $J_X^{\prime\alpha}$ are transformed into each other under the action of parity, which is a unitary operator. 
As a result, integrating over the phase space of $|X_s\rangle$ for any $q^2$ value, 
and summing (averaging) over the spin configurations of both $|X_s\rangle$ 
and $|H_b\rangle$, leads to the same 
coefficients in Eq.~(\ref{eq:rx_dGdq})
for $\left|C^{\ell}_L\right|^2$ and  $\left| C^{\ell\prime}_L\right|^2$.
Moreover, the positivity of the squared matrix element implies 
\be
| \etanot (q^2) | \leq 2\, ,
\qquad 
\etaseven (q^2)  >0\,.
\label{eq:etanotbound}
\ee

Given the definition of $R_X$ in  Eq.~(\ref{eq:RX}),  it is convenient to define 
the following $q^2$-integrated hadronic parameters:
\begin{align}\label{eq:hadr_param}
    F_X^{\ell} &=\int_{q_{\rm min}^2}^{q_{\rm max}^2} f_X^{\ell}(q^2) dq^2, \nonumber \\
    \ril &= \frac{1}{F_X^{\ell}}\int_{q_{\rm min}^2}^{q_{\rm max}^2} f_X^{\ell}(q^2) \etai (q^2) dq^2 .
\end{align}
The normalization factor  $f_X^{\ell}(q^2)$ depends on the lepton mass via kinematic effects, 
which are sizeable only close to the endpoint (i.e.~for $q^2 \to 4m_\ell^2$). If the $q^2$ range of the measurement extends well above the di-lepton mass threshold, 
the lepton mass dependence is safely neglected and we can set
\be
F_X^{\mu}=F_X^{e} \equiv F_X\, , \qquad  \ril \equiv \ri\, . 
\ee
In this limit the overall normalization factor  drops out  in $R_X$ and 
the same hadronic parameters appear in both numerator and denominator:
\begin{widetext}
\be
R_X = \frac{   \Big\{ \left|C^{\mu}_L\right|^2 + \left| C^{\mu\prime}_L\right|^2  + \Re\left[  \rnot C^{\mu*}_L C^{\mu\prime}_L 
+ C^*_7 \left(  \rseven   C_7 + \rone C^{\mu}_L   +\rtwo  C^{\mu\prime}_L   \right) \right] \Big\} + \Big\{ L\to R \Big\} 
   }{      \Big\{ \left|C^{e}_L\right|^2 + \left| C^{e\prime}_L\right|^2  + \Re\left[  \rnot C^{e*}_L C^{e\prime}_L 
+ C^*_7 \left(  \rseven   C_7 + \rone C^{e}_L   +\rtwo  C^{e\prime}_L   \right) \right] \Big\} + \Big\{ L \to R \Big\} }\, .
\label{eq:RXfull}
\ee
\end{widetext}
This implies that in the SM, and in all models where the Wilson coefficients are lepton universal, 
$R_X\approx 1$
up to  corrections due to QED and/or residual kinematic effects which are at most of O(1\%)~\cite{Bordone:2016gaq,Isidori:2020acz}.

The key observation of the present 
work is that 
$R_X$ retains a significant discriminating power with respect to NP models 
even in the absence of a precise knowledge of the hadronic parameters, i.e.~even when treating the
 $\ri$ as  nuisance parameters.  This statement emerges quite clearly by the following two observations:
 \begin{itemize}
 \item{}  Sizeable deviations of $R_X$ from unity can only be attributed to non-universal 
Wilson coefficients, i.e.~$|R_X - 1| \not =0$ only if $|\Delta C_i| \not =0 $ for some  $i$, where
\be 
\Delta C_i  = C^\mu_i - C^e_i\, , \qquad i={L, L^\prime, R, R^\prime}\, . 
\ee
 \item{}   Other observables constrain NP effects to be a small perturbation over the SM:
 this implies that large NP effects in $R_X$ can arise only by non-vanishing $\Delta C_i$ 
 interfering with the SM amplitude.  The  latter has a peculiar structure, 
 \bea
&&  |C^{\rm SM}_L|  = O(10) \gg   |C^{\rm SM}_7|, \,  |C^{\rm SM}_R|,   \no\\
 && |C^{\ell\prime}_{L,R}|^{\rm SM}=0\, ,
 \eea
 hence only a very limited set of NP amplitudes can lead to $ |R_X - 1|  \gg 0$.
 \end{itemize}
 
These two observations become evident when
linearising the theoretical expression of $R_X$ with respect to the $\Delta C_i$ and neglecting the interference of 
$\Delta C_i$ with suppressed SM amplitudes.
In this limit we obtain 
\be
R_X  - 1 \approx \frac{ \Re\left(   2 \frac{\Delta C_L}{ C^{\rm SM}_L}   +  \rnot    \frac{\Delta C^\prime_{L}}{ C^{\rm SM}_L}   
\right)   }{ 1 +  \rseven \left| \frac{ C^{\rm SM}_7 }{ C^{\rm SM}_L } \right|^2  +  \Re\left[ \rone  \frac{ C^{\rm SM}_7 }{ C^{\rm SM}_L }  \right]}\,.
\label{eq:RXapp}
\ee
As can be seen, only two types of NP effects can lead to a sizeable 
deviation of $R_X$ from one: a lepton non-universal shift in either $C_L^\ell$ or $C_L^{\ell\prime}$.
Note also that the only hadronic parameter with direct impact on the extraction of 
NP constraints from $R_X$ is $\etanot$, which is bounded  by Eq.~(\ref{eq:etanotbound}). The $\etaseven$ and 
$\etaone$ parameters have a minor role: they control the {\em dilution} of the
LFU violation in the rate due to the lepton-universal contribution  by $\cO_7$.
Finally, the effect of $\etatwo$ is always subleading.

\begin{figure}[t]
\centering
\includegraphics[width=0.45\textwidth]{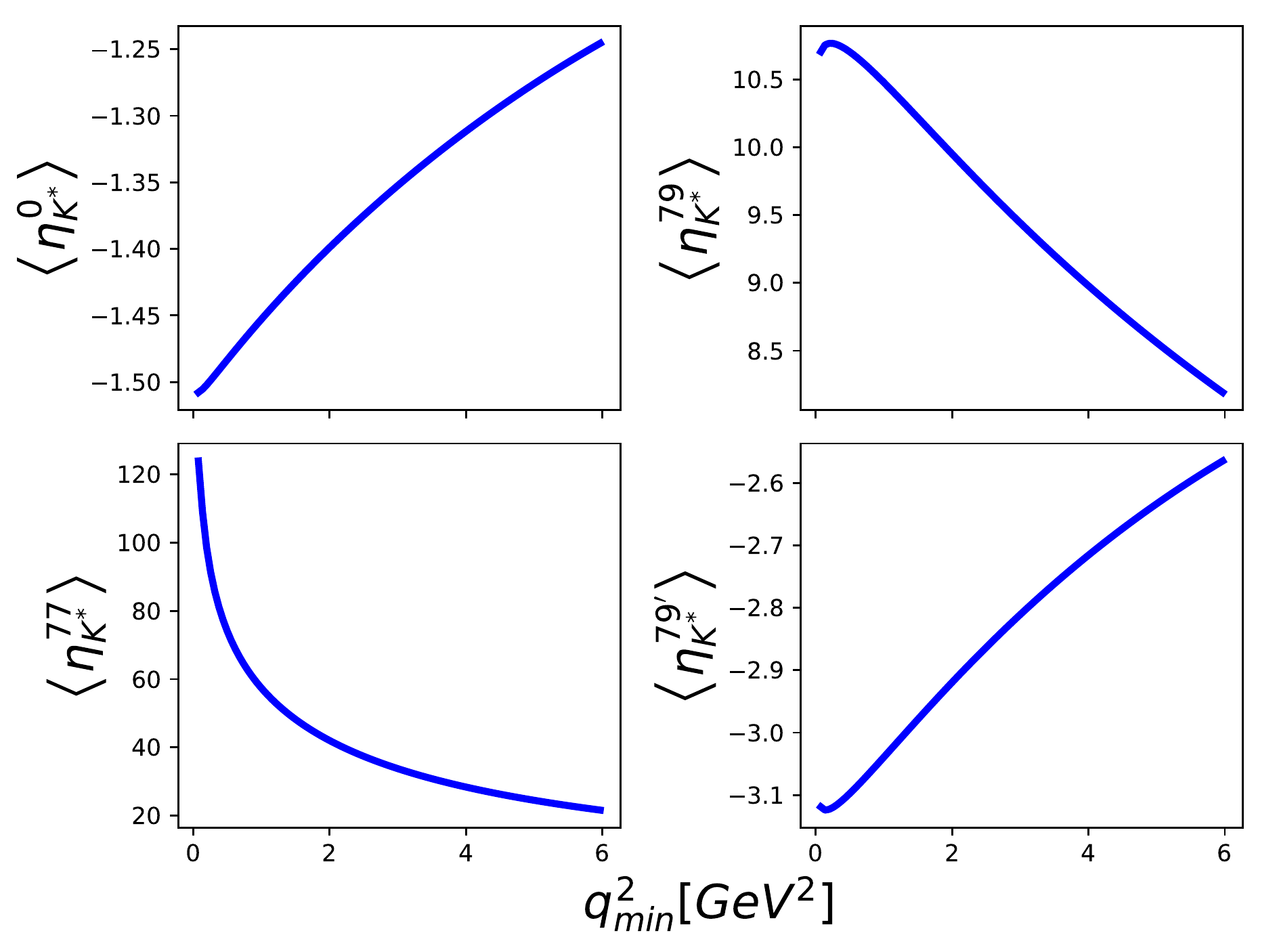}
\caption{Integrated hadronic parameters $\ri$, defined in Eq.~(\ref{eq:hadr_param}), 
extracted from $B^0\to K^{*0}(\to K \pi)\mu^+\mu^-$  as a function of $q_{\rm min}^{2}$, setting $q^{2}_{\rm max}=6~{\rm GeV}^{2}$.}
\label{fig:hadron_paramters2}
\end{figure}

The approximate expression in Eq.~(\ref{eq:RXapp}) is shown for illustrative purposes only, in the following
numerical analysis we use  the complete expression 
in Eq.~(\ref{eq:RXfull}),
treating all the  $\ri$ as  nuisance parameters. 
In order to define a range for  the $\rseveni$,  we use a channel where we are able to compute the values of the $\ri$ parameters explicitly 
and where the impact of the dipole operator is maximal, namely the $B^0\to K^{*}(892)^{0} \ell^+\ell^-$ decay.
In this mode, characterised by a spin-one final state, the dipole operator is maximally enhanced by the $q^2 \to 0$ pole.
 In multi-body channels, such as $B^0\to K^{+} \pi^{-} \ell^+\ell^-$ and $B^{+}\to K^{+} \pi^{-}\pi^{+} \ell^+\ell^-$, 
 with a sizeable $S$-wave component of the hadronic final state, we expect a significantly smaller contribution 
 of $\cO_7$ to the total decay rate. 
The values for  the $\ri$ for this channel 
as a function of  $q_{\rm min}^2$, setting $q^2_{\rm max}=6~{\rm GeV}^2$, are shown Fig.~\ref{fig:hadron_paramters2}. 
The corresponding ranges for the hadronic parameters used in the numerical analysis 
are shown in Table~\ref{tab:rpkpars}.\footnote{Note that the large value of $\rseven$ is largely compensated by 
the smallness of $C_7$: even if $\rseven =O(100)$, $\rseven |C_7|^2 = O(10) \ll |C^{\rm SM}_L |^2 = O(100)$.}

We conclude this section  with a few observations related to the theoretical expression of 
$R_X$:
\begin{itemize}
\item{}
In Eq.~(\ref{eq:rx_dGdq}) we ignored the contribution to the rate of four-quark operators. In the $q^2$ region far from the narrow charmonia, dominated by perturbative contributions, their effect is small and {\em cannot} induce a violation of LFU. Similarly to $\cO_7$, four-quark operators can only induce a dilution of the LFU contribution.  Their effect can indeed be described as a $q^2$ dependence modification of  coefficient $C_9$, which would leave  Eq.~(\ref{eq:RXapp}) unchanged up to an irrelevant shift in $C_L^{\rm SM}$.
\item{}
The parameter $\etanot$ weights the relative
contribution of vector and axial currents in the hadronic transition, and is maximal for hadronic final states with well-defined parity. In the 
$B\to K$ case, where only the vector current contributes, $\eta_K^0=2$; in the 
$B\to K^*$ case, which is dominated by the axial-current contribution, $ -2 < \eta_{K^*}^0 < -1$; in the fully inclusive case $\etanot \approx 0$.
\item{}
As pointed first in~\cite{Hiller:2014ula}
(see also~\cite{Fuentes-Martin:2019mun}),
in the motivated class of NP models where the 
lepton non-universal amplitudes have a pure left-handed structure, 
the value of $R_X$ is expected to be the same
for any $B\to X_s \ell^+\ell^-$ transition:
\be 
\left. ( R_X  -1)\right|_{ \Delta C_L \not=0 } 
\approx \left. ( R_K -1 )\right|_{ \Delta C_L \not=0} \,.  
\ee
\end{itemize}

\begin{table}[t]
  \centering
    \begin{tabular}{ l  c   }
      \hline
      Parameter & Limits  \\
      \hline \\ [-7pt]
$\rnot$ &  [-2,2]  \\[3pt]
$\rone$ &  [-12,12]   \\[3pt]
$\rtwo$ &  [-4,4]  \\[3pt]
$\rsevenpk$ &  [0,120]   \\[3pt]
$\rsevenkpipi$ &  [0,60]   \\[2pt]
      \hline
    \end{tabular}
  \caption{Limits placed on the hadronic nuisance parameters. A larger range is used for $\rsevenpk$ compared to $\rsevenkpipi$ due to the wide $q^{2}$ range used in the experimental measurement~\cite{Aaij:2019wad}.}
  \label{tab:rpkpars}
\end{table}

\section{Global combination of current measurements}\label{sec:stat}

In this section we present a 
combination of \mbox{\bsll} measurements 
following the procedure described in Ref.~\cite{Isidori:2021vtc}. We include the following three sets of observables: i)~the LFU ratios $R_K$~\cite{Aaij:2021vac},  $R_{K^*}$~\cite{Aaij:2017vbb} and \rpk~\cite{LHCb:2019efc}, ii)~the branching ratio for the rare dilepton mode $B_s^{0}\to \mu^+\mu^-$~\cite{CMS:2014xfa,Aaboud:2018mst,Sirunyan:2019xdu,Beneke:2019slt} and, iii)~the normalised angular distribution in $B^{0}\to K^{*0}\mu^+\mu^-$ decays~\cite{Aaij:2020nrf,Aaij:2020ruw}.
In the case of $R_K$ and $R_{K^*}$, where the structure of the 
hadronic matrix elements is well understood, we use the standard
theoretical expressions in terms of Wilson coefficients and form factors 
using the Flavio package~\cite{Straub:2018kue}.
The  ratio \rpk\ is described by means of Eq.~(\ref{eq:RXfull}).

As discussed in Ref.~\cite{Isidori:2021vtc}, 
we employ a highly generic NP hypothesis and a highly conservative approach towards hadronic uncertainties. We generate pseudo-experiments according to the SM, fluctuating the measurements according to their experimental uncertainties, and calculate the likelihood ratio between the NP and SM hypotheses. The distribution of the likelihood ratio is then used to calculate the p-value of a fit to data. Long-distance charm contributions are treated by allowing for a lepton universal shift of $\cO^\ell_{9}$ in the SM definition.

The lepton universality ratio \rpk~has been measured by the LHCb collaboration to be consistent with unity in the $q^{2}$ region $0.1<q^{2} < 6.0$ GeV$^{2}/c^{4}$~\cite{LHCb:2019efc}. We include it in the combination by means of Eq.~(\ref{eq:RXfull}), using the limits reported in Table~\ref{tab:rpkpars} for the hadronic parameters.
In fact, preliminary results on the differential branching fraction intervals of the dimuon invariant mass further confirms the smaller contribution of 
$\cO_7$ to the total rate~\cite{Lisovskyi:2699822}, if compared to 
the benchmark $B^0\to K^{*}(892)^{0} \mu^+\mu^-$ decay~\cite{LHCb:2016ykl}.
As four nuisance parameters are included with only one measurement, degeneracies in the likelihood can occur due to multiple solutions. To counteract this, 
loose Gaussian constraints, whose width is the size of the physical ranges, are placed on each parameter to ensure the likelihood has a well-defined minimum. The exact value of these ranges has a very small effect on the numerical results.

\begin{figure}[t]
\centering
\includegraphics[width=0.45\textwidth]{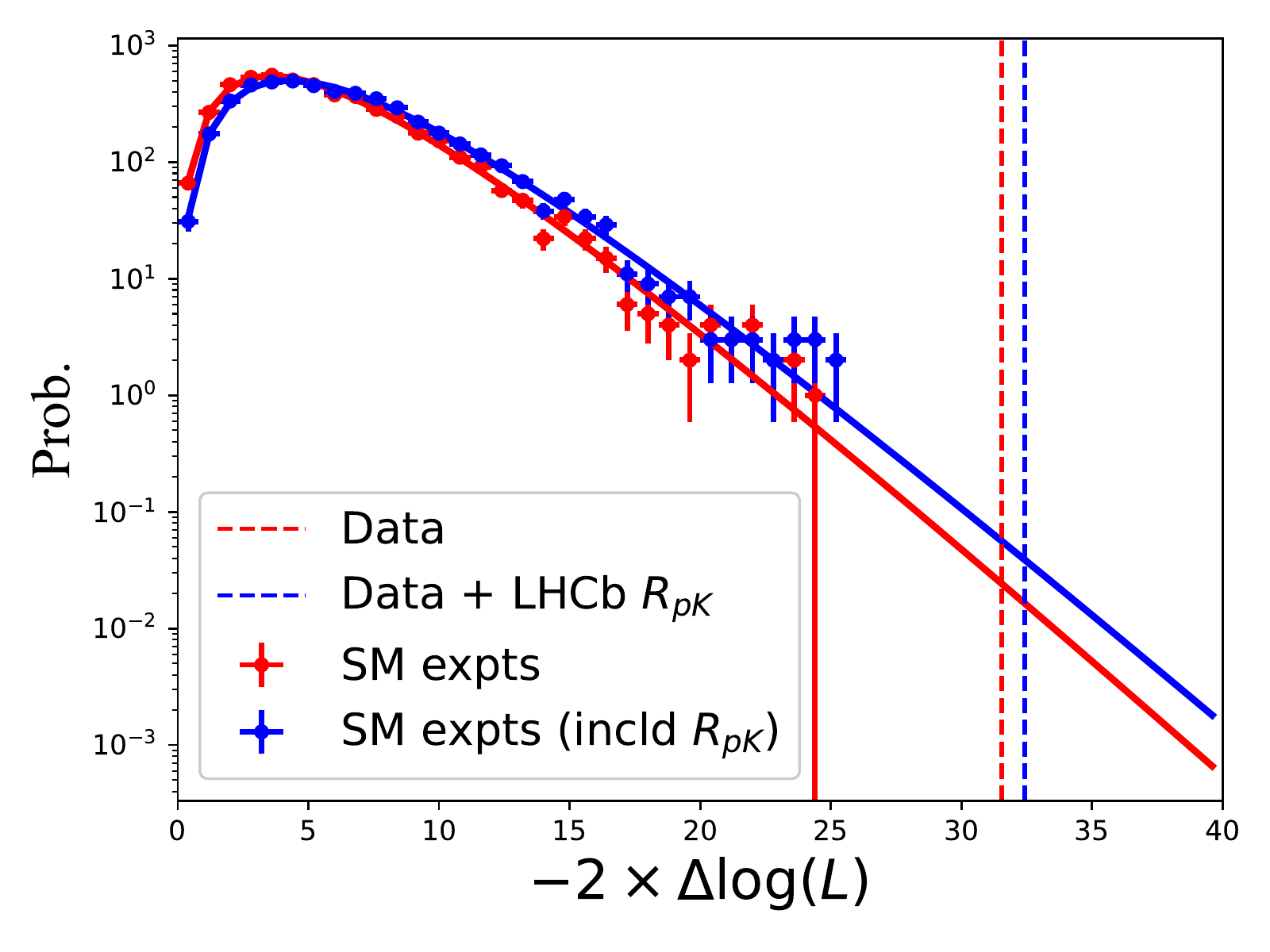}
\caption{
Distribution of the likelihood ratio for pseudo-experiments under the SM hypothesis along with the value obtained from data. Results are shown under the same conditions as in Ref.~\cite{Isidori:2021vtc} and also when the measurements of $R_{pK}$ is included.}


\label{fig:global_lhcbrpk}
\end{figure}


The distribution of the likelihood ratio for the SM pseudoexperiemnts is shown in Fig~\ref{fig:global_lhcbrpk}, along with the value obtained from data. The inclusion of the measurement of $R_{pK}$ increases the effective degrees of freedom by 0.6 units. This increase represents the uncertainty on the 
$\ri$ which allows for potentially different NP sensitivity compared to the existing $R_K$ and $R_{K^{*}}$ ratios. Compared with the results from Ref.~\cite{Isidori:2021vtc}, we observe a small reduction in significance, from $4.3\sigma$ to $4.2\sigma$ when including the observable \rpk. This is due to the fact that the value of \rpk\ is not perfectly consistent with the other LFU ratios and the hadronic uncertainties 
allow to accommodate deviations from the SM
amplitude in other directions, 
within a general NP hypothesis.


Using the same approach we test  
the specific hypothesis of a violation of lepton universality, considering all \Rx\ ratios measured so far, i.e.~including $R_K$, $R_{K^*}$ and \rpk,
and ignoring all other observables.  
This results in a local significance of $4.1\sigma$ for
the hypothesis of a LFU violation, 
which is very close to the global significance of NP in 
\bsll decays. This small variation in the significance can be understood as follows: the 
analysis of LFU observables has a smaller trial factor compared 
to the generic NP analysis; 
however, with present data, this effect is compensated by the lack of 
inclusion in the fit 
of $\cB(B_s\to \mu^+\mu^-)$~\cite{Sirunyan:2019xdu,Aaboud:2018mst,CMS:2014xfa,LHCb:2021awg,LHCb:2021vsc},
which enhances the significance in the generic NP case.

\section{Impact of future measurements}\label{sec:prospect}

In addition to assessing the significance with the current measurements, we calculate the expected gain in discovery potential by using this approach with other non-exclusive \Rx\ measurements that can be performed at LHCb in the near future. To this end, we estimate the experimental sensitivity of these ratios and include the hypothetical measurements in a fit with the current measurements. 

We estimate the experimental sensitivity of three modes with the full run I and run II dataset of 9$\rm fb^{-1}$ for the following ratios:
\begin{align}\label{eq:hadr_param}
    R_{pK} &=\frac{{\cal B}(\Lambda_b^{0}\to p K^{-} \mu^+\mu^-)}{{\cal B}(\Lambda_b^{0}\to p K^{-} e^+e^-)}, ~ \nonumber \\
    R_{K\pi\pi} &=\frac{{\cal B}(B^{+}\to K^{+}\pi^{-}\pi^{+} \mu^+\mu^-)}{{\cal B}(B^{+}\to K^{+}\pi^{-}\pi^{+} e^+ e^-)}, \nonumber \\
    R_{K\pi} &=\frac{{\cal B}(B^{0}\to K^{+}\pi^{+} \mu^+\mu^-)}{{\cal B}(B^{0}\to K^{+}\pi^{+} e^+ e^-)}, ~ \nonumber
\end{align}
where for the \rkpi\ case, the $K^{+} \pi^{-}$ invariant mass is required to be above 1~GeV to separate it from the comparatively well understood \kstar resonance.

The sensitivity for non-exclusive \Rx\ measurements depends primarily on the precision of the electron mode. Given the ratio \rpk\ has already been measured, the precision can easily be predicted assuming it scales with luminosity, resulting in a precision of $12.2\%$. As the decays $B^{+}\to K^{+}\pi^{-}\pi^{+} e^{+}e^{-}$ and $B^{0}\to K^{+}\pi^{-} e^{+}e^{-}$ have yet to be observed, their yields are extrapolated from the corresponding muonic decay modes from Refs.~\cite{LHCb:2016eyu,LHCb:2014osj}, by scaling with luminosity and 
the centre-of-mass energy. These muon yields are compared to the corresponding yield in the $R_{K^{*}}$ measurement~\cite{LHCb:2017avl} to scale the resulting precision of the LFU ratio. A statistical uncertainty on \rkpi\ and \rkpipi\ of $7.7\,\%$ and $13.5\,\%$ is expected for the full run I-II datasets in the range of $1.1< q^{2} < 6.0$ GeV$^{2}/c^{4}$. 
The estimated uncertainty on \rkpi\ turns out to be 
comparable with that of $R_{K^{*}}$, as can be expected given 
there are many significant contributions above the \kstar resonance~\cite{LHCb:2016eyu,Lu:2011jm}.

\begin{figure}[t]
\centering
\includegraphics[width=0.45\textwidth]{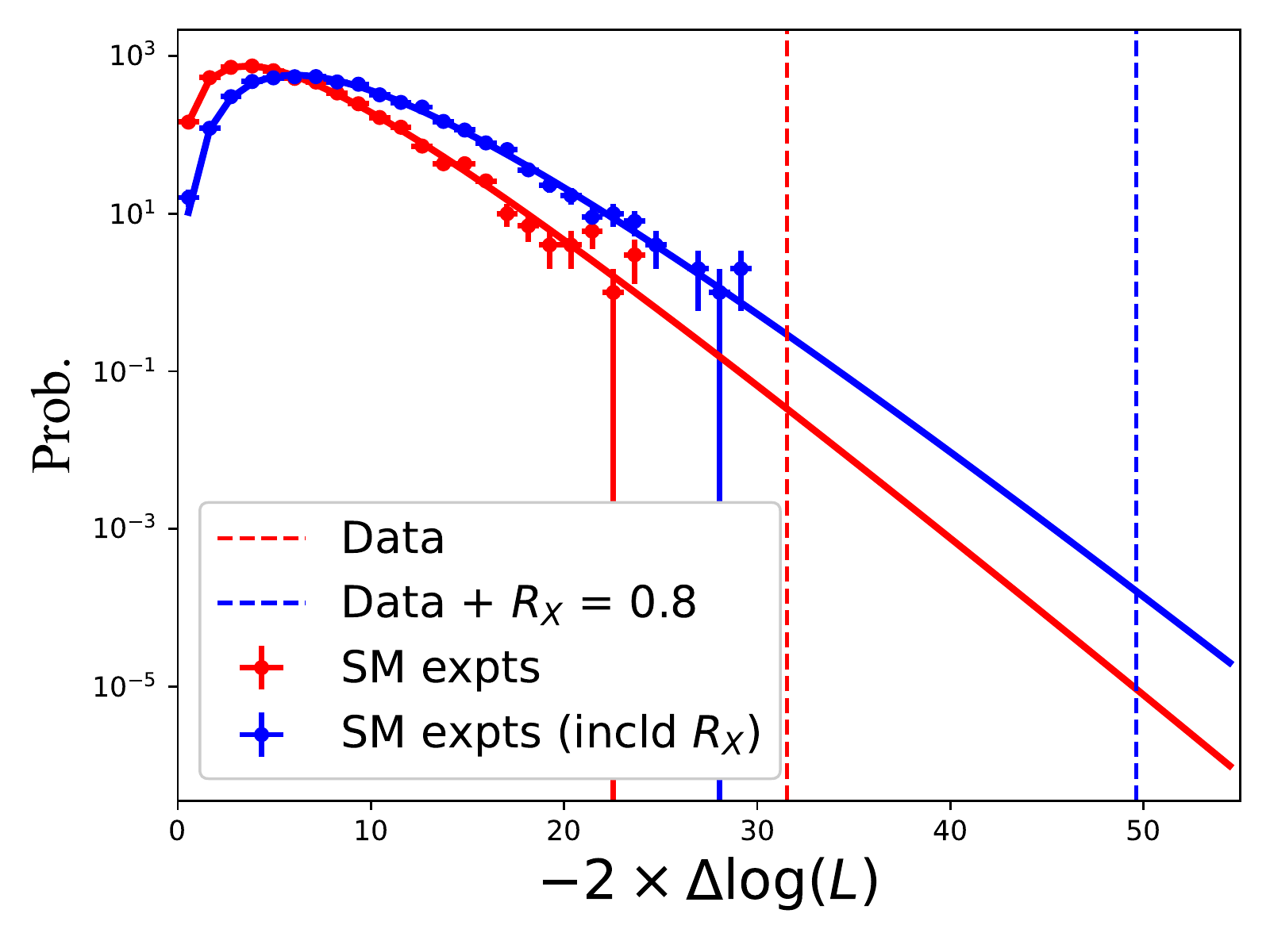}
\includegraphics[width=0.45\textwidth]{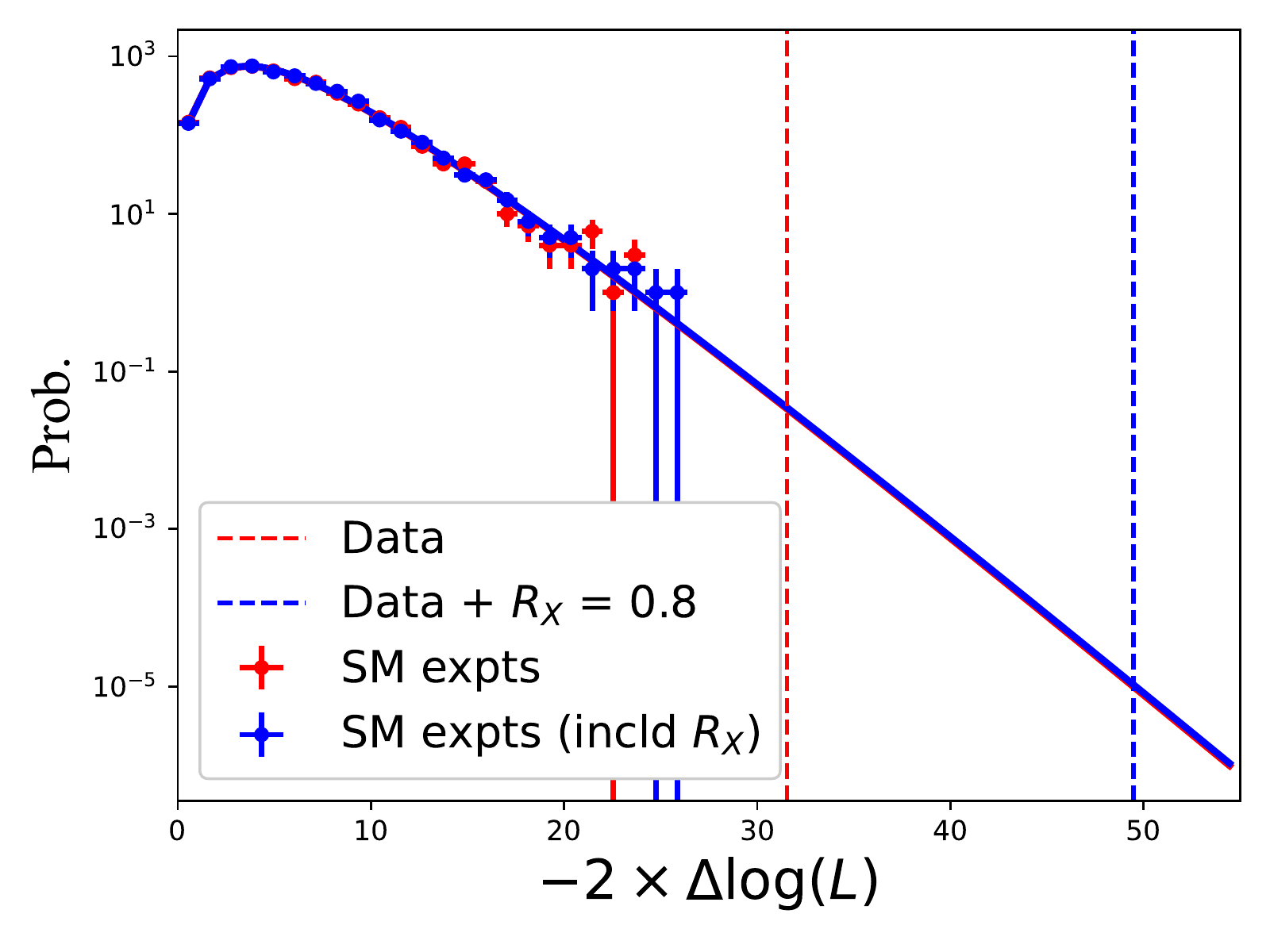}
\caption{
Distribution of the likelihood ratio for pseudo-experiments under the SM hypothesis along with the value obtained from data. The distribution is overlaid with a scenario including hypothetical non-exclusive $R_X$ measurements along with their expected sensitivities (blue). An azimov dataset~\cite{Cowan:2010js} is used to estimate the expectation value for the significance. 
}
\label{fig:global_smallrconstrained}
\end{figure}




Information on the differential branching fraction in intervals of the dimuon invariant mass can provide insights on the underlying dynamics of the non-exclusive hadronic system, which allows us to check the limits of the $\ri$ parameters. For instance, for the $B^0\to K^{*0}\mu^+\mu^-$ decay a relative increase of the differential branching fraction between the $0.1< q^{2} < 0.98$ and $1.1 < q^{2} < 6.0$ GeV$^{2}/c^{4}$ regions by a factor of three is reported in Ref.~\cite{LHCb:2016ykl}. Similar inspection can be performed for the non-exclusive channels and are found to be at the order of $2.0$ and $2.6$ for the $K\pi$ and $K\pi\pi$ hadronic systems, 
respectively~\cite{LHCb:2016eyu,LHCb:2014osj}.
As a result, this confirms the conclusion of Sec~\ref{sec:theory} that the limits obtained for the $\ri$ parameters involving the \kstar resonance can be used as a proxy for these channels.  \\

The impact of these future measurements is examined by repeating the procedure from the previous section introducing two benchmark points
common to all non-exclusive LFU ratios: $R_{X} = 1.0$ (SM) 
and $R_{X} = 0.8$ (NP). 
The latter is chosen  
being broadly consistent with current global fits. 
Figure~\ref{fig:global_smallrconstrained} (top) shows the distribution of the likelihood ratio when including these new $R_{X}$ observables under the NP hypothesis.
A large increase in the significance from $4.3\,\sigma$ to $5.4\,\sigma$ when including the $R_{X}$ observables is seen. 
If the new measurements are set to the SM prediction of $R_{X}=1.0$, a reduction to $3.8\,\sigma$ can be expected. These measurements can therefore have a large impact on the clarification of lepton universality violation in \bsll decays.

\begin{table}[t]
  \centering
    \begin{tabular}{ l  c   }
      \hline
      \multicolumn{1}{c}{Scenario} & NP Significance  \\
      \hline \\ [-7pt]
Current data  &  4.3 $\sigma$   \\[3pt]
Current data + $R_X = 0.8$ &  5.4 $\sigma$   \\[3pt]
Current data + $R_X = 1.0$ &  3.8 $\sigma$   \\[2pt]
      \hline
    \end{tabular}
  \caption{Change of the significance of the new-physics hypothesis in \bsll decays 
  adding hypothetical measurements of  \rpk, \rkpi, and \rkpipi, with full run I and run II
  statistics, under two different hypotheses for the central values.}
  \label{tab:sig}
\end{table}

In order to investigate the dependence of the significance with respect to the freedom given to the hadronic parameters, we have repeated 
the fit fixing the $\ri$ to their central values. The result is also shown in Fig.~\ref{fig:global_smallrconstrained} (bottom).
As expected, in this case the additional measurements do not increase the effective degrees of freedom in the system. The exact knowledge of all hadronic nuisance parameters would lead to a significance of $5.9\sigma$, i.e.~an increase in significance of $0.5\sigma$ compared to when they are treated as nuisance parameters. 
This relatively small increase provides an a posteriori   
confirmation that they play a 
minor role in the fit.
Finally, we also decrease the limits allowed for $\rseven$ to 60, which would be appropriate if the \rpk\ ratio were measured setting $q^2_{\rm min}$ above 1~GeV$^{2}$. A negligible difference in discovery potential is seen, which indicates that the exact kinematic range is not crucial for the subsequent interpretation.

\section{Conclusions}\label{sec:conclusions}

In summary, we have introduced a method to include any LFU ratio in global fits by treating the hadronic uncertainties as nuisance parameters. This method is not designed to replace the existing theoretical description of $R_K$ or $R_{K^{*}}$, where we can take advantage of a detailed knowledge of 
all the components of the transition amplitudes.
It is conceived for interpreting LFU ratios where 
we lack precise information about the 
underlying hadronic dynamics. 

To demonstrate the method, we have updated the global fit of Ref.~\cite{Isidori:2021vtc} to include the LHCb measurement of \rpk. With current data, we find 
that  \rpk\ has a marginal effect on the
global 
significance of new physics in \bsll decays. 
However, when extrapolating to the full LHCb dataset, and including also hypothetical measurements of \rkpi\ and \rkpipi, 
we find that the increase in the significance can be large. 

In this paper we concentrated on the three 
non-exclusive LFU ratios  
which are more promising from the experimental 
point of view. However, the method proposed here 
can be extended to include many other channels,
such as $B^+\to K^{+}K^{-}K^{+}\ell^{+}\ell^{-}$.
An interesting experimental feature of some of the
non-exclusive channels is that, due to the large invariant mass of the hadronic systems,
they 
suffer much less from partially 
reconstructed backgrounds compared to the {\em golden modes}
$B^0\to K^{*0}\ell^+\ell^-$ and $B^+\to K^+\ell^+\ell^-$. 
This additional experimental advantage reduces the risk of hypothetical mis-modelling of backgrounds, which right now are among the leading systematic uncertainties in the LFU measurements.
The inclusion of the non-exclusive $R_X$
using the method proposed here will therefore 
not only increase the new-physics sensitivity 
from a pure statistical point of view, but also enhance the redundancy of the experimental results.

\acknowledgements
This work was inspired by questions asked in the Rare Decays Working Group of LHCb, we acknowledge the role of the lively and intellectually stimulating environment of this working group in attracting our attention on this problem. 
This project has received funding from the European Research Council (ERC) via the European Union's Horizon 2020 research and innovation programme under grant agreement 833280 (FLAY), and from the Swiss National Science Foundation (SNF) under contracts 182622 and 174182.

\bibliography{letter_PLB}

\end{document}